\documentclass[11pt,a4paper]{article}
\pdfoutput=1 
\usepackage[sc]{mathpazo}
\usepackage[compact]{titlesec}

\usepackage{amssymb,amsthm,amsmath,mathtools}
\usepackage{thmtools,thm-restate}
\usepackage[round]{natbib}
\usepackage{pifont}

\usepackage[linesnumbered,lined,boxed,ruled,vlined]{algorithm2e}
\DontPrintSemicolon
\SetKwInOut{Parameters}{Parameters}
\SetKwComment{Comment}{$\triangleright$\ }{}
\SetAlFnt{\small}
\SetAlCapFnt{\small}
\SetAlCapNameFnt{\small}
\SetAlCapHSkip{0pt}
\IncMargin{-\parindent}

\usepackage[usenames,svgnames,xcdraw,table]{xcolor}
\definecolor{DarkGreen}{rgb}{0.1,0.5,0.1}
\usepackage[backref=page]{hyperref}
\hypersetup{
	colorlinks=true,
	linkcolor=red,
	urlcolor=DarkGreen,
	citecolor=blue
}
\renewcommand*{\backref}[1]{}
\renewcommand*{\backrefalt}[4]{%
    \ifcase #1 (Not cited.)%
    \or        (Cited on page~#2)%
    \else      (Cited on pages~#2)%
    \fi}
\usepackage[capitalise,noabbrev]{cleveref}
\Crefname{property}{Property}{Properties}
\Crefname{theorem}{Theorem}{Theorems}
\Crefname{example}{Example}{Examples}
\Crefname{table}{Table}{Tables}
\Crefname{algorithm}{Algorithm}{Algorithms}

\usepackage{cancel}
\usepackage{tabularx}
\usepackage{nicefrac}
\usepackage{enumerate}
\usepackage{etoolbox}
\usepackage[margin=1in]{geometry}
\usepackage[T1]{fontenc}
\usepackage{url}
\usepackage[utf8]{inputenc}
\usepackage{graphicx}
\usepackage{booktabs}
\usepackage{comment}

\usepackage{mathrsfs,amsfonts,dsfont,authblk}
\usepackage{array,multirow,graphicx,bigdelim}

\usepackage{pgf,tikz}
\usepackage{tikz-dimline}
\usepackage{tikz-cd}
\usetikzlibrary{fit,calc,math,shapes,decorations.text,arrows,decorations.markings,decorations.pathmorphing,shapes.geometric,positioning,decorations.pathreplacing}
\tikzset{snake it/.style={decorate, decoration=snake}}

\colorlet{mygray}{gray!40}

\makeatletter
\let\oldnl\nl
\newcommand{\nonl}{\renewcommand{\nl}{\let\nl\oldnl}}
\makeatother

  {\list{}{\leftmargin=#1\rightmargin=#1}\item[]}%
  {\endlist}

\usepackage[labelfont={normalfont,bf},textfont=it]{caption}
\usepackage{subcaption}

\newtheorem{theorem}{Theorem}
\newtheorem{lemma}{Lemma}

\newtheorem{observation}{Observation}
\newtheorem{corollary}{Corollary}
\theoremstyle{definition}

\newtheorem{examplex}{Example}

\newenvironment{example}{\pushQED{\qed}\examplex}{\popQED\endexamplex}
\theoremstyle{remark}

\newtheorem{definition}{Definition}

\usepackage{flushend}
\usepackage[utf8]{inputenc}
\usepackage[inline]{enumitem}
\Crefname{claim}{Claim}{Claims}

\allowdisplaybreaks

\newcommand{\SPC}{\textrm{\textup{SPC}}}
\newcommand{\NCC}{\textrm{\textup{NCC}}}
\newcommand{\USM}{\textrm{\textup{USM}}}
\newcommand{\MP}{\textrm{\textup{MaxProp}}}
\newcommand{\MR}{\textrm{\textup{MaxRou}}}
\newcommand{\DA}{\textrm{\textup{DA}}}
\newcommand{\mMP}{\textrm{\textup{m-MaxProp}}}
\newcommand{\mMR}{\textrm{\textup{m-MaxRou}}}
\newcommand{\wMP}{\textrm{\textup{w-MaxProp}}}
\newcommand{\wMR}{\textrm{\textup{w-MaxRou}}}

\title{A Gale-Shapley View of {\em Unique} Stable Marriages}
\author[1]{Kartik Gokhale}
\author[1]{Amit Kumar Mallik}
\author[1,2]{Ankit Kumar Misra}
\author[1]{Swaprava Nath}

\affil[1]{\small Indian Institute of Technology Bombay, Mumbai, India\\\texttt{kartikgokhale2000@gmail.com,19D070007@iitb.ac.in,swaprava@iitb.ac.in} }
\affil[2]{\small University of California, Los Angeles\\ \texttt{ankitkmisra@ucla.edu} }
\date{}

\begin{document}

\maketitle

\begin{abstract}
    Stable marriage of a two-sided market with unit demand is a classic problem that arises in many real-world scenarios. In addition, a unique stable marriage in this market simplifies a host of downstream desiderata. In this paper, we explore a {\em new} set of sufficient conditions for {\em unique stable matching} (\USM{}) under this setup. Unlike other approaches that also address this question using the structure of preference profiles, we use an algorithmic viewpoint and investigate if this question can be answered using the lens of the {\em deferred acceptance} (\DA{}) algorithm~\citep{gale1962college} without actually running the algorithm. Our results yield a set of sufficient conditions for \USM{} (viz., \MP{} and \MR{}) and show that these are disjoint from the previously known sufficiency conditions like {\em sequential preference} and {\em no crossing}. We provide a characterization of \MP{} that makes it efficiently verifiable (without using \DA{}).
    These results give a more detailed view of the sub-structures of the \USM{} class.
\end{abstract}

\section{Introduction}
\label{sec:intro}

The {\em stable marriage problem} considers a two-sided market where agents of each side (e.g., men) are assumed to have a linear preference over the other side (e.g., women) and matches are one-to-one, i.e., each agent has a single demand. Stability asks for a pairing between these agents such that there does not exist any pair of a man and a woman who would like to abandon the current matching and mutually prefer a marriage among themselves. \citet{gale1962college} proved that such a stable matching always exists and is obtained via a computationally simple algorithm called {\em deferred acceptance} (\DA{}). However, there could be multiple stable matchings and it raises questions on which one to pick. The stable matching problem is very well studied in the literature and several useful results exist related to \DA{} and its variants. For instance, the questions regarding the maximum~\citep{karlin2018simply} or average number of stable matchings~\citep{pittel1989average}, complexity of counting stable marriages~\citep{irving1986complexity}, matching with incomplete lists \citep{iwama2002stable}, indifferences \citep{manlove2002structure}, heterogeneous jobs and workers \citep{crawford1981job}, and many more, have already been investigated. See \citet{iwama2008survey,manlove2013algorithmics} for a comprehensive survey on the stable matching problem and \citet{roth2008deferred} for a survey of the \DA{}-type algorithms.

In this context, uniqueness of stable matching \citep{eeckhout2000uniqueness,clark2006uniqueness} has a very important place. First, since the actual pairings of men and women are a stable matching based on their \textit{reported} preferences, a normative goal is to ensure that it is indeed their {\em actual} preferences, i.e., the stable matching algorithm is {\em strategyproof}. However, it is known that \DA{} is not strategyproof for a non-proposer~\citep{gale1985ms} unless there is a unique stable matching. 
Though a unique stable matching is not sufficient for strategyproofness~\citep{roth1989two} except in the incomplete information setup~\citep{ehlers2007incomplete}, it is a property from which further structures of strategyproofness can be obtained. 
We define the class of preference profiles where the set of stable matchings is a singleton as {\em unique stable matching} (\USM{}) in this paper. 

The second reason why \USM{} is desirable is the anti-symmetry of the preferences of men and women over the stable matchings. It is known that between two different stable matchings $\mu_1$ and $\mu_2$, if $\mu_1$ is at least as preferred as $\mu_2$ by all men, then $\mu_2$ must be at least as preferred as $\mu_1$ by all women, i.e., men and women have exactly opposite preferences over the stable matchings~\citep{gale1985some}. 
Hence, finding a stable matching that is {\em unbiased} to any side of the market is often challenging. A considerable amount of research effort has been put to find a {\em fair} compromise between the two extremes (see, e.g., \citet{klaus2006procedurally,tziavelis2020fair,brilliantova2022fair}).
However, the question of bias also does not appear in the \USM{} class since there is exactly one stable matching.

Finally, unique stable matchings have appeared in many real-world matching markets, e.g., in the US National Resident Matching Program \citep{roth1999redesign}, Boston school choice \citep{pathak2008leveling}, online dating \citep{hitsch2010matching}, and the Indian marriage market \citep{banerjee2013marry}.

In this paper, we aim to understand the internal structure of the \USM{} class using a \DA{} algorithmic lens.

\begin{figure*}[t!]
    \definecolor{qqttcc}{rgb}{0,0.2,0.8}
    \definecolor{zzqqww}{rgb}{0.6,0,0.4}
    \definecolor{ttttff}{rgb}{0.2,0.2,1}
    \definecolor{ffqqqq}{rgb}{1,0,0}
    \definecolor{qqqqff}{rgb}{0,0,1}
    \definecolor{fftttt}{rgb}{1,0.2,0.2}
     \centering
     \begin{subfigure}[b]{0.45\textwidth}
         \centering
\scalebox{0.70}{
    \begin{tikzpicture}[line cap=round,line join=round,>=triangle 45,x=1.0cm,y=1.0cm]
    \clip(0.16,-4.04) rectangle (11.84,4.06);
    \draw [fill=black,fill opacity=0] (2.86,-0.04) circle (2.17cm);
    \draw [fill=black,fill opacity=0] (2.88,-0.94) circle (1.17cm);
    \draw [line width=1.2pt,dash pattern=on 4pt off 4pt,color=ffqqqq,fill=ffqqqq,fill opacity=0.25] (7.66,1.82) circle (1.73cm);
    \draw [line width=1.2pt,dash pattern=on 4pt off 4pt,color=ttttff,fill=ttttff,fill opacity=0.25] (7.68,-1.88) circle (1.73cm);
    \draw [dash pattern=on 4pt off 4pt,color=zzqqww,fill=zzqqww,fill opacity=0.25] (8,2) circle (0.91cm);
    \draw [line width=1.2pt,dash pattern=on 4pt off 4pt,color=qqttcc,fill=qqttcc,fill opacity=0.25] (8,-2) circle (0.91cm);
    \draw [rotate around={0:(6,0)}] (6,0) ellipse (5.65cm and 3.99cm);
    \draw (3.66,3.26) node[anchor=north west] {\Huge \USM{}};
    \draw (6.62,1.0) node[anchor=north west] {\Large \mMP{}};
    \draw (7.1,2.28) node[anchor=north west] {\Large \mMR};
    \draw (2.34,1.12) node[anchor=north west] {\Large \SPC{}};
    \draw (2.22,-0.94) node[anchor=north west] {\Large \NCC{}};
    \draw (6.62,-0.36) node[anchor=north west] {\Large \wMP{}};
    \draw (7.1,-1.6) node[anchor=north west] {\Large \wMR{}};
    \end{tikzpicture}
}
         \caption{$n \geqslant 3$}
         \label{fig:ngreater3}
     \end{subfigure}
     \hfill
     \begin{subfigure}[b]{0.45\textwidth}
         \centering
\scalebox{0.85}{
    \begin{tikzpicture}[line cap=round,line join=round,>=triangle 45,x=1.0cm,y=1.0cm]
    \clip(1.72,-2.4) rectangle (10.64,3.22);
    \draw (3.26,2.22) node[anchor=north west] {\LARGE \USM{} = \SPC{} = \NCC{}};
    \draw (6.86,0.14) node[anchor=north west] {\parbox{2.37 cm}{\mMP{} \newline= \mMR{}}};
    \draw [rotate around={0.97:(7.46,-0.3)},line width=1.6pt,dash pattern=on 4pt off 4pt,color=fftttt,fill=fftttt,fill opacity=0.2] (7.46,-0.3) ellipse (1.93cm and 1.53cm);
    \draw [rotate around={-0.61:(6.03,0.37)},line width=1.2pt,dash pattern=on 1pt off 1pt on 1pt off 4pt,fill=black,fill opacity=0.1] (6.03,0.37) ellipse (3.91cm and 2.7cm);
    \draw [rotate around={0.97:(5,-0.44)},line width=1.6pt,dash pattern=on 4pt off 4pt,color=qqqqff,fill=qqqqff,fill opacity=0.2] (5,-0.44) ellipse (1.93cm and 1.53cm);
    \draw (3.14,0.14) node[anchor=north west] {\parbox{2.34 cm}{\wMP{} \newline= \wMR{}}};
    \end{tikzpicture}
}
         \caption{$n = 2$}
         \label{fig:nequal2}
     \end{subfigure}
        \caption{The above two figures illustrate the sub-structures of the \USM{} class for $n \geqslant 3$ and $n=2$ respectively. However, the gap between \MP{} and \MR{} is empty for $n=3$ (\Cref{lemma:mp-mr-n=3}) and non-empty for $n\geqslant 4$. The dashed lines and the shaded regions denote the new sub-structures of \USM{} that are contributions of this paper. We also characterize the class \MP{} and provide the complexity of verification. In \Cref{fig:nequal2}, the fact \USM{} = \SPC{} was known from \citet{eeckhout2000uniqueness}. We provide a direct proof of this fact.}
        \label{fig:contrib_illus}
\end{figure*}

\subsection{Our contributions}
\label{sec:contrib}

The main contributions of this paper are as follows (illustrated graphically in \Cref{fig:contrib_illus}).
\begin{itemize}
    \item We view the \USM{} problem using the number of proposals and rounds in the classic Gale-Shapley \DA{} algorithm, and introduce two new conditions, men-proposing \MP{} or \mMP{} and men-proposing \MR{} or \mMR{} (similarly \wMP{} and \wMR{} for the women-proposing versions), defined w.r.t.\ men (women)-proposing \DA{}. These properties identify those preference profiles where a men (women)-proposing DA algorithm takes maximum possible number of proposals or rounds respectively. We show the mutual relationship of these two properties in \Cref{thm:maxr-implies-maxp} when the number of men (or women) $|M| (= |W|) = n \geqslant 3$. We show that each of these conditions is {\em sufficient} for \USM{} (\Cref{thm:maxp-implies-usm}).
    \item We characterize the class \MP{} in \Cref{thm:necessary-sufficient-maxp} and show that these conditions are efficiently verifiable (\Cref{thm:complexity}) without requiring any appeal to \DA{}.
    \item Prominent existing sufficient conditions for \USM{}, the {\em sequential preference condition} (\SPC{}~\citep{eeckhout2000uniqueness}) and the {\em no crossing condition} (\NCC{}~\citep{clark2006uniqueness}), are disjoint from the new sufficient conditions proposed in this paper for $n \geqslant 3$ (\Cref{thm:maxp-spc-disjoint}). Hence, our results make the internal sub-structure of the \USM{} class outside \NCC{} and \SPC{} clearer. In particular, the men and women proposing versions of \MP{} class turns out to be disjoint as well (\Cref{thm:mmp-wmp-disjoint}).
    \item When $n=2$, we show that the classes \mMP{} and \mMR{} (similarly \wMP{} and \wMR{}) coincide (\Cref{thm:maxp-implies-maxr-n=2}) and so do \SPC{} and \NCC{} (\Cref{thm:spc-implies-ncc-n=2}). Also, \mMP{} and \wMP{} are contained within \SPC{} for $n=2$ (\Cref{thm:maxp-implies-spc-n=2}). We also provide a direct proof of the fact that for $n=2$, \USM{} and \SPC{} are equivalent (\Cref{thm:usm-implies-spc-n=2}), a result originally proved by \citet{eeckhout2000uniqueness}. However, we also point out an inconsistency in the claim of \SPC{} being necessary for \USM{} for $n=3$~\citep{eeckhout2000uniqueness} through \Cref{ex:usm-not-maxp-not-spc}.
    Interestingly, for $n=2$, the classes \mMP{} and \wMP{} have an overlap and we characterize it in \Cref{thm:partial-overlap-n=2}.
\end{itemize}

\subsection{Related works}
\label{sec:lit}

Several works have focused on finding sufficient conditions for \USM{}, the two most well-known of these being the sequential preference condition~\citep{eeckhout2000uniqueness} and the no crossing condition~\citep{clark2006uniqueness}. Others include the co-ranking condition~\citep{legros2010co}, the acyclicity condition~\citep{romero2013acyclicity}, the universality condition~\citep{holzman2014matching}, oriented preferences~\citep{reny2021simple}, aligned preferences~\citep{niederle2009decentralized}, uniqueness consistency~\citep{karpov}, $\alpha$-reducibility~\citep{alcalde1994exchange}, and iterative $\alpha$-reducibility~\citep{park}. These results provide structural insight into the types of preference profiles that lead to uniqueness in stable matchings. Refer to~\Cref{sec:sota} for a detailed discussion on previously known sufficient conditions for USM.

It is noteworthy that nearly all of the sufficient conditions listed above are either restrictions or generalizations of the sequential preference condition (\SPC{}) or the no crossing condition (\NCC{}). In contrast, our work unveils two sufficient conditions that have no overlap at all with \SPC{} or \NCC{}, as we prove later.

Finding a necessary condition has also been investigated, and there are two prominent approaches. The first one uses the idea of $\alpha$-reducibility, proposed originally by \citet{alcalde1994exchange}. A marriage problem satisfies $\alpha$-reducibility if every sub-population of men and women has a {\em fixed pair} (a pair of man and woman who prefer each other the most). \citet{clark2006uniqueness} shows that this condition is both necessary and sufficient for the existence of a unique stable matching in {\em every sub-population} of men and women. If USM is considered only for the full population, $\alpha$-reducibility is sufficient {\em but not necessary}.

A different approach to this problem uses the idea of {\em acyclicity}, originally proposed by \citet{chung2000existence}. Acyclicity implies that if the agents point to their most preferred partners, then the resulting directed graph should not have any directed cycle. While \citet{romero2013acyclicity} show that it is a sufficient condition for \USM{}, the necessity condition using this method is explored recently by \citet{gutin2021unique}. \citet{gutin2021unique} use the acyclicity on a reduced graph that they define as the {\em normal form}. The idea of normal form is used for submatching markets by \citet{irving1986complexity}, and \citet{balinski1997stable}. \citet{gutin2021unique} claim that the difficulty in finding a necessary condition for \USM{} in these approaches was that the acyclicity property was being used on the complete preference profile, while the entire preference profile may not be relevant for a unique stable matching. Using the idea of normal form, they prune the preferences where an agent can never match with certain partners in any stable matching. This acyclicity on a normal form turns out to be necessary and sufficient for \USM{}~\citep{gutin2021unique}.

Our approach differs considerably from those discussed above, in the way our conditions are defined. Instead of looking at the \USM{} class through the preference structures of the players, we view it using the \DA{} algorithm and its execution over a profile. Our results consider the maximum number of proposals made by the agents and the number of rounds in \DA{}, and provide the extra structures that yield a clearer view of the space between the currently known sufficient conditions and the \USM{} class (\Cref{fig:contrib_illus}). It shows that certain properties of an algorithm can also help clarify the structure of \USM{}, without even running that algorithm.

\section{Preliminaries}
\label{sec:prelim}

Consider a two-sided unit-demand matching market, where the two sides are represented, WLOG, by men and women respectively. The agents of each side are expressed as two equi-cardinal finite sets, denoted by $M$ and $W$, $|M|=|W|=n$, respectively. The sets share no common agents, i.e., $M \cap W = \emptyset$. All men have strict preferences over all women and vice versa. Individual preferences, denoted $\succ_i$ for agent $i$, are assumed to be complete, transitive, and anti-symmetric. The notation $m_i \succ_{w_k} m_j$ denotes $w_k\in W$ prefers $m_i \in M$ over $m_j \in M$, and similarly, $w_i \succ_{m_k} w_j$ denotes $m_k \in M$ prefers $w_i\in W$ over $w_j\in W$. The preference profile is denoted by $\succ := \{\succ_i: i \in M \cup W\}$. The set of all complete, transitive, and anti-symmetric preference profiles in this setup is denoted by $\mathcal{P}$. A {\em matching} and several other definitions in this setting follow~\citet{gale1962college}.

\begin{definition}[Matching]
    \label{def:matching}
    A matching in $\succ$ is a mapping $\mu$ from $M \cup W$ to itself such that for every man $m \in M, \mu(m) \in W$, for every woman $w \in W, \mu(w) \in M$, and for every $m, w \in M \cup W$, $\mu(m) = w$ if and only if $\mu(w) = m$.
\end{definition}

The above definition says that each man is matched to exactly one woman and vice-versa.
To define stability of a matching, we need the definition of {\em blocking pair} as given below.

\begin{definition}[Blocking Pair]
    \label{def:blocking}
    A pair $(m,w), m \in M, w \in W$ is a blocking pair of a matching $\mu$ in $\succ$ if $m \succ_w \mu(w)$ and $w \succ_m \mu(m)$.
\end{definition}
Informally, the above definition means that the pair $(m,w)$ both prefer each other over their currently matched partners.
This leads to the definition of stable matching as follows.

\begin{definition}[Stable Matching]
    \label{def:stable}
    A matching $\mu$ in $\succ$ is {\em stable} if it does not have any blocking pair.
\end{definition}

\citet{gale1962college} showed that for any preference profile $\succ$, a stable matching always exists and can be found via the {\em deferred acceptance} (\DA{}) algorithm. The working principle of this algorithm is the following. The algorithm comes in two versions based on whether the men or the women are the proposers. In every round of the men-proposing \DA{} algorithm, each unmatched man proposes to his favorite woman that has not rejected him already. Each woman, in that round, receives the proposals and {\em tentatively} accepts the most favorite man that has proposed to her and rejects the rest. The rejected men go to the next round and repeat this activity. The algorithm stops when no man remains unmatched.
A formal representation is given in \Cref{algo:da}.

\begin{algorithm}
\caption{(Men-proposing) Deferred Acceptance (\DA)}
\label{algo:da}
    \KwIn{men $M=\{m_1,\dots,m_n\}$, women $W=\{w_1,\dots,w_n\}$, and preferences $\succ=\{\succ_i: i \in M \cup W\}$}
    \KwOut{a stable matching}
    \For{$i\in M\cup W$}{
        $\mu(i) \gets \emptyset$
    }
    \While{$\exists m\in M$ such that $\mu(m) = \emptyset$}{
        \tcp*[l]{This is a round}
        \For{$m_i\in M$ such that $\mu(m_i) = \emptyset$}{
            \tcp*[l]{This is a proposal}
            $w \gets$ highest woman in $\succ_{m_i}$ to whom $m_i$ has not proposed yet\\
            \If{$\exists m_j\in M$ such that $\mu(m_j)=w$}{
                \If{$m_i\succ_w m_j$}{
                    $\mu(m_i) \gets w$, $\mu(w) \gets m_i$\\
                    $\mu(m_j) \gets \emptyset$\\
                }
            }
            \Else{
                $\mu(m_i) \gets w$, $\mu(w) \gets m_i$
            }
        }
    }
    \Return{$\mu$}
\end{algorithm}

\paragraph{} The following two facts about DA will be used frequently throughout this paper.

\begin{restatable}{fact}{singleproposal}
\label{fact:single-proposal}
    In the men-proposing \DA{} algorithm, there exists a woman $w \in W$ who receives exactly one proposal.
\end{restatable}
\begin{proof}
    We prove this by contradiction. Suppose there exists a preference profile $\succ$ where each woman gets at least two proposals. Consider the last round $r$ of men-proposing \DA{} on $\succ$. Since every woman gets at least two proposals in total, every woman must have received at least one proposal before round $r$. This means every woman, and hence every man, is matched at the end of round $r-1$, causing termination of \DA{}. This is a contradiction to round $r$ being the last round of \DA{}. Thus, the fact is proved.
\end{proof}

\begin{restatable}{fact}{boundachievable}
\label{fact:bound-achievable}
    In the men-proposing \DA{} algorithm, the maximum possible number of proposals is $n^2-n+1$, and the maximum possible number of rounds is $n^2-2n+2$.
    Both the bounds are {\em achievable}, i.e., there exists a preference profile $\succ \in \mathcal{P}$ where the above numbers are attained.
\end{restatable}
\begin{proof}
    By \Cref{fact:single-proposal}, there is a woman who receives exactly one proposal. WLOG, say $w_n$ is one such woman. The other women $w_1,\ldots,w_{n-1}$ can receive up to a maximum of $n$ proposals, one from each man. This suggests an upper bound of $n(n-1)+1 = n^2-n+1$ on the number of proposals in men-proposing \DA{}.
    
    Moreover, all men make proposals in the first round, so the first round must consist of $n$ proposals, whereas all the remaining rounds must have at least one proposal. Together with the above upper bound on the number of proposals, this implies an upper bound of $(n^2-n+1)-n+1 = n^2-2n+2$ on the number of rounds.

    In the following preference profile, these upper bounds are achieved.
    \begin{itemize}
        \item For $i \in \{1,\ldots,n-1\}$, $m_i$ has preference $w_i \succ w_{i+1} \succ \dots \succ w_{n-1} \succ w_1 \succ w_2 \succ \dots \succ w_{i-1} \succ w_n$.
        \item $m_n$ has preference $w_1 \succ w_2 \succ \dots \succ w_n$.
        \item For $j\in\{1,\ldots,n\}$, $w_j$ has preference $m_{j+1} \succ m_{j+2} \succ \dots \succ m_n \succ m_1 \succ m_2 \succ \dots \succ m_j$.
    \end{itemize}
    With the above preferences, $w_n$ gets exactly one proposal, and all the men $m_1,\ldots,m_n$ cycle through women $w_1,\ldots,w_{n-1}$ one by one until the final assignment of $m_i \leftrightarrow w_{i-1}, i = 2, \ldots, n$ and $m_n \leftrightarrow w_n$. As argued while getting the expressions of the upper bounds on the number of proposals and rounds, this structure is where the $(n-1)$ women except $w_n$ receive $n$ proposals each and $w_n$ receives only one proposal. Also, this structure has $n$ proposals in the first round and each subsequent round has exactly one proposal made. This is the recipe for getting $n^2-n+1$ proposals and $n^2-2n+2$ rounds. So, clearly this profile achieves the upper bound.
\end{proof}

Although DA always converges to a stable matching, it is also known that men-proposing \DA{} and women-proposing \DA{} converge to men and women optimal stable matchings respectively, which could be quite different. There is a hierarchy among the stable matchings from the men and women points of view as given by the following result.

\begin{theorem}[\citet{gale1985some}]
    \label{thm:gale-sotomayor-inverse}
    If for any two distinct stable matchings $\mu_1$ and $\mu_2$ in $\succ$, if each man finds $\mu_1$ at least as preferred as $\mu_2$, then every woman will find $\mu_2$ at least as preferred as $\mu_1$.
\end{theorem}

The subclass of $\mathcal{P}$ where the set of stable matchings is a singleton is defined as the {\em unique stable matching} (\USM{}) class. In \USM{}, the men and women proposing \DA{} reach the same stable matching. Because of the various satisfactory properties exhibited by this class as discussed in \Cref{sec:intro}, there had been various attempts to characterize the structures of the preference profiles in \USM{}. In the following section, we introduce two prominent sufficient conditions for \USM{}. 

\paragraph{Remark.} There are certain necessity results of \USM{} as well, using ideas like $\alpha$-reducibility~\citep{clark2006uniqueness} and {\em acyclicity} using a {\em normal form} of the preferences~\citep{gutin2021unique}. However, in this paper, our objective is to view it from a \DA{} algorithmic perspective and we discuss how our results can be applicable without running \DA{} and even in domains with partial preferences (\Cref{sec:gap}).

\section{Current State-of-the-art Sufficient Conditions}
\label{sec:sota}

Although there have been various sufficient conditions proposed for \USM{} \citep[e.g.]{romero2013acyclicity,gusfield1989parametric,reny2021simple}, the {\em sequential preference condition} (\SPC{}, \citep{eeckhout2000uniqueness}) and  {\em no crossing condition} (\NCC{}, \citep{clark2006uniqueness}) provide a deeper structural view of the preference profiles of the agents that gives rise to \USM{}.

\begin{definition}[Sequential Preference Condition]
    \label{def:spc}
    A preference profile $\succ$ satisfies {\em sequential preference condition} (\SPC{}) if there exists an ordering of men, $m_1, m_2, \ldots, m_n$, and women, $w_1, w_2, \ldots, w_n$, such that
    \begin{enumerate}
        \item man $m_i$ prefers $w_i$ over $w_{i+1}, w_{i+2}, \ldots, w_n$, and
        \item woman $w_i$ prefers $m_i$ over $m_{i+1}, m_{i+2}, \ldots, m_n$.
    \end{enumerate}
\end{definition}
\citet{eeckhout2000uniqueness} showed that \SPC{} is sufficient for uniqueness of stable matching. It is, however, not necessary for $n \geqslant 3$ as we show in the example below. 

\begin{example}[\USM{} but not \SPC{}]
    \label{ex:not-spc}
    Consider the following preference profile.
    \[
    \left(
    \begin{array}{cc}
       m_1: & w_2 \succ w_1 \succ w_3 \\
       m_2: & w_1 \succ w_2 \succ w_3 \\
       m_3: & w_1 \succ w_2 \succ w_3
    \end{array};
    \
    \begin{array}{cc}
       w_1: & m_1 \succ m_2 \succ m_3 \\
       w_2: & m_2 \succ m_3 \succ m_1 \\
       w_3: & m_3 \succ m_2 \succ m_1
    \end{array}
    \right)
    \]
    This does not satisfy \SPC{}, since \SPC{} needs at least one pair of man and woman that rank each other at the top. However, the men-proposing \DA{} yields the matching where $m_i$ is matched with $w_i$, $i = 1,2,3$, which is the men-optimal matching. However, in this case, that is the women-optimal as well since each woman gets her top preference. By \Cref{thm:gale-sotomayor-inverse}, this profile has an unique stable matching, i.e., it belongs to \USM{}.
\end{example}

Later, \citet{clark2006uniqueness} defined the following refinement that implies \SPC{}.

\begin{definition}[No Crossing Condition]
    \label{def:ncc}
    A preference profile $\succ$ satisfies {\em no crossing condition} (\NCC{}) if there exists an ordering $(m_1,m_2,\ldots,m_n)$ of $M$ and an ordering $(w_1,w_2,\ldots,w_n)$ of $W$, such that if $i<j$ and $k<l$, then 
    \begin{enumerate}
        \item \label{cond:ncc-1} $w_l \succ_{m_i} w_k \Rightarrow w_l \succ_{m_j} w_k$, and
        \item \label{cond:ncc-2} $m_j \succ_{w_k} m_i \Rightarrow m_j \succ_{w_l} m_i$.
    \end{enumerate}
\end{definition}
This condition implies that if the men and women are lined up in that given order and any pair of men (or women) are asked to point to his (or her) favorite partner among a pair of potential partners, their pointers cannot cross each other. Though \NCC{} implies \SPC{}, the converse is not true for $n \geqslant 3$~\citep{clark2006uniqueness}. 
%
%
These sufficient conditions for $n \geqslant 3$ are shown on the LHS of \Cref{fig:ngreater3}. \SPC{} and \NCC{}, however, become identical with \USM{} for $n=2$ as we discuss later.

Following the discovery of \SPC{} and \NCC{}, various other sufficient conditions for \USM{} have been proposed, the majority of them being either restrictions or generalizations of the former two conditions. For instance, the co-ranking condition of~\citet{legros2010co} and the universality condition of~\citet{holzman2014matching} are both contained within \NCC{}, while the uniqueness consistency condition of~\citet{karpov} relaxes \SPC{}, and the oriented preferences of~\citet{reny2021simple} and aligned preferences of~\citet{niederle2009decentralized} generalize \SPC{} to many-to-one markets. Even the $\alpha$-reducibility condition of~\citet{alcalde1994exchange} lies between \NCC{} and \SPC{}, and its iterative version from~\citet{park} coincides with \SPC{} itself.

Since the \SPC{} and \NCC{} conditions clearly take center stage in the current state-of-the-art in sufficient conditions for \USM{}, we limit the comparison of our new conditions to only these two. \Cref{ex:not-spc} shows that there exists unexplored space in $\USM{} \cap \overline{\SPC{}}$. We provide additional structure to that space in this paper.

\section{Our Results}
\label{sec:results}

This paper considers the \USM{} problem from the \DA{} perspective. We first define two new conditions that we prove to be sufficient for \USM{}. The definitions deal with the number of proposals women get in men-proposing \DA{} and the number of rounds involved. In the rest of the paper, WLOG, we use men-proposing \DA{} whenever we consider \DA{}. However, analogous definitions and results hold for a symmetrically opposite women-proposing version as well.
\Cref{fact:bound-achievable} prompts us to define the following two classes of preferences.

\subsection{MaxProposals and MaxRounds}
\label{sec:maxR-maxP}

These two classes of preferences are defined as follows.

\begin{definition}[\MP{} and \MR{}]
    \label{def:maxr-maxp}
    A preference profile $\succ$ satisfies
    \begin{enumerate}
        \item \MP{}, if the proposers make $n^2-n+1$ proposals in \DA{}, and
        \item \MR{}, if the proposing process in \DA{} happens for $n^2-2n+2$ rounds.
    \end{enumerate}
\end{definition}
Note that, the above two classes are critically dependent on the proposing side. We will denote the classes where the maximum number of proposals (and rounds) are coming from the men-proposing \DA{} as \mMP{} (and \mMR{}) respectively. The women-proposing versions of the classes will be denoted as \wMP{} and \wMR{} respectively. In the rest of the paper, WLOG, we will imply the men-proposing versions of \MP{} and \MR{} respectively when we refer to them and prove their properties. The results for the women-proposing versions are identical and are skipped. However, in \Cref{sec:men-women-maxp-disjoint}, we show that the classes \mMP{} and \wMP{} are disjoint for $n \geqslant 3$. Interestingly, these two classes partially overlap for $n=2$, and we discuss it in \Cref{sec:n=2}.
Our first result shows the relationship between the classes \MP{} and \MR{}. 

\begin{theorem}
    \label{thm:maxr-implies-maxp}
    If a preference profile $\succ$ satisfies \MR{}, then $\succ$ also satisfies \MP{}.
\end{theorem}

\begin{proof}
    WLOG, assume men-proposing \DA{} in this case.
    Suppose a preference profile $\succ$ satisfies \MR{}. This implies that if we run the men-proposing \DA{} algorithm, it would take $n^2-2n+2$ rounds to terminate. We make the following observations directly from the algorithm.
    \begin{itemize}
        \item The first round involves $n$ proposals as nobody is matched yet, i.e., each man makes a proposal.
        \item Each round (except the last one) must see at least one man getting rejected, else the termination criterion of the algorithm is met, and thus, every round (except the first one) has at least one proposal.
    \end{itemize}
    Hence, the total number of proposals in $\succ$ is $\geqslant n + n^2-2n+1 = n^2-n+1$.
    By \Cref{fact:bound-achievable}, we know that the number of proposals is at most $n^2-n+1$. Hence, the number of proposals in $\succ$ must be $= n^2-n+1$. Therefore $\succ$ satisfies \MP{}.
\end{proof}

The converse of the above theorem is not true for $n \geqslant 4$ as the following example shows.

\begin{example}[$\MP{}$ but not $\MR{}$ for $n \geqslant 4$]
\label{ex:maxp-not-maxr}
    Consider the following preference profile involving four men and four women.
    \begin{center}
       \[\left(
       \begin{array}{cc}
       m_1: & w_1 \succ w_2 \succ w_3 \succ w_4 \\
       m_2: & w_3 \succ w_2 \succ w_1 \succ w_4 \\
       m_3: & w_3 \succ w_1 \succ w_2 \succ w_4 \\
       m_4: & w_1 \succ w_2 \succ w_3 \succ w_4
    \end{array};
    \
    \begin{array}{cc}
       w_1: & m_2 \succ m_3 \succ m_4 \succ m_1 \\
       w_2: & m_3 \succ m_4 \succ m_1 \succ m_2 \\
       w_3: & m_4 \succ m_1 \succ m_2 \succ m_3 \\
       w_4: & m_1 \succ m_2 \succ m_3 \succ m_4 
    \end{array}
    \right)\]
    \end{center}
    In this example, two men ($m_1$ and $m_3$) get rejected in the first round of \DA{}. Both these men propose in the next round and it is easy to check that the number of proposals for this profile is $n^2-n+1 = 13$. However, \MR{} requires one proposal per round (thereby increases the number of rounds to the maximum). But here we have two proposals in round 2, which fails \MR{}.
\end{example}

\paragraph{Remark.}
It turns out that for $n=3$, \MP{} implies \MR{} and thus both conditions are equivalent. The proof can be found in \Cref{app:mp=mr-n=3}.

We now state an important lemma which will be used in the following sections to prove several properties of \MP{}. The result gives structure to the proposals observed in profiles satisfying \MP{}.

\begin{lemma}
\label{lemma:structure-maxp}
    WLOG, let $w_n$ be the woman who receives exactly one proposal in men-proposing \DA{} on $\succ$. If $\succ \in \MP{}$, then all men $m \in M$ propose to all women in $W \setminus \{w_n\}$.
\end{lemma}

\begin{proof}
    Since $\succ \in \MP{}$, we have $n^2 - n + 1$ proposals. Since $w_n$ receives exactly one proposal, the other $n-1$ women receive a total of $n^2 - n$ proposals. No woman can receive more than $n$ proposals (since there are $n$ men). Hence, the only way $n-1$ women can receive $n^2-n$ proposals is if each woman in $W \setminus \{w_n\}$ receives $n$ proposals. Thus, all $m \in M$ must propose to all $w \in W \setminus \{w_n\}$.
\end{proof}


    
        
Notice that, if a woman receives proposals from all men, she is always assigned to her most preferred man according to the men-proposing \DA{}. Hence, the following corollary is immediate from the lemma above.
\begin{corollary}
    \label{cor:maxp-women-get-top}
    If $\succ \in \MP{}$, all women except the one who gets exactly one proposal, get matched with their most preferred men. Formally, if $w_n$ is the woman who gets exactly one proposal, then for all $i\in\{1,\dots,n-1\}$, $\mu(w_i) \succ_{w_i} m_j$ or $\mu(w_i) = m_j$ for all $j \in [n]$.
\end{corollary}

\subsection{\MP{} implies \USM{}}
\label{sec:mp-to-usm}

In this section, we prove one of the major results of this paper that provides a new sufficient condition of \USM{}.

\begin{theorem}
    \label{thm:maxp-implies-usm}
    If a preference profile $\succ$ satisfies \MP{}, then $\succ$ is in \USM{}.
\end{theorem}

\begin{proof}
    Suppose a preference profile $\succ$ satisfies \MP{}. We show that the (men-optimal) output of men-proposed \DA{} algorithm (say $\mu$) is also women-optimal. It is also known that the men(women)-optimal stable match is unique~\citep{roth1992two}. Then, together with \Cref{thm:gale-sotomayor-inverse}, $\mu$ would be the unique stable matching.
    
    Let $w_n$ be the woman who receives exactly one proposal. By \Cref{cor:maxp-women-get-top}, all other women are matched with their first preferences. 

    Suppose, there is another stable matching $\mu' \neq \mu$ on the same profile $\succ$, which is more preferable than $\mu$ for women. Then, for all $i \in [n]$, either $\mu'(w_i) \succ_{w_i} \mu(w_i)$ or $\mu'(w_i) = \mu(w_i)$, and for some $j \in [n], \mu'(w_j) \succ_{w_j} \mu(w_j)$.
    
    However, $w_1, w_2, \ldots, w_{n-1}$ are already matched to their first preferences by $\mu$. So, $\mu'(w_i) = \mu(w_i)$ for $i=1,\ldots,n-1$, and $\mu(w_n)$ has to be the only man remaining who has to be matched to $w_n$ even in $\mu'$. Hence, $\mu=\mu'$, which is a contradiction. Thus, $\mu$ is women-optimal, and is the unique stable matching.
\end{proof}

However, the converse of the previous theorem is not true. The following example shows that \MP{} is not necessary for \USM{}. In fact, this example does not satisfy \SPC{} either.

\begin{example}[\USM{} but neither \MP{} nor \SPC{}]
    \label{ex:usm-not-maxp-not-spc}
    Consider the following preference profile.
    \[
    \left(
    \begin{array}{cc}
       m_1: & w_1 \succ w_3 \succ w_2 \\
       m_2: & w_2 \succ w_1 \succ w_3 \\
       m_3: & w_1 \succ w_2 \succ w_3
    \end{array};
    \
    \begin{array}{cc}
       w_1: & m_2 \succ m_1 \succ m_3 \\
       w_2: & m_3 \succ m_1 \succ m_2 \\
       w_3: & m_1 \succ m_2 \succ m_3
    \end{array}
    \right)
    \]
    Since there is no pair of man and woman $(m,w)$ that prefers each other the highest, it is not \SPC{}. The men-proposing \DA{} takes $6$ proposals, while the maximum number of proposals is $3^2 - 3 + 1 = 7$. Hence, this profile does not satisfy \MP{}. However, the men-optimal matching (obtained via men-proposing \DA{}) results in all women receiving their most preferred men, which is women-optimal as well. Therefore, this profile belongs to \USM{}.
\end{example}

From \Cref{thm:maxr-implies-maxp,thm:maxp-implies-usm}, the following corollary is immediate.

\begin{corollary}
    \label{cor:maxr-implies-usm}
    If a preference profile $\succ$ satisfies \MR{}, then $\succ$ is in \USM{}.
\end{corollary}

\section{A Characterization of \MP{}}
\label{sec:gap}

In this section, we find the conditions of the preference profiles that are necessary and sufficient for \MP{}. We also show that these certifications of belonging to \MP{} can be done efficiently without involving the \DA{} algorithm.
We begin with a few structural properties of \MP{}.

\begin{lemma}
\label{lemma:struct-maxp-1}
    If a preference profile $\succ \in \mathcal{P}$ satisfies \MP{}, then there must be a woman $w \in W$ who is the {\em least preferred} woman for each $m \in M$.
\end{lemma}

\begin{proof}
    We prove this result via contradiction.
    WLOG, suppose woman $w_n$ is the woman who receives exactly one proposal (by \Cref{fact:single-proposal}) when men-proposing \DA{} is run on $\succ$. Suppose there is a man $m_i$ who does not have $w_n$ as his last preference. Let $m_i$ prefer $w_n$ over some woman $w_j$, $j \neq n$. Then by \Cref{lemma:structure-maxp} (as $\succ$ satisfies \MP{}), $m_i$ must propose to $w_j$, and since he prefers $w_n$ over $w_j$, he must propose to $w_n$ before $w_j$. But, $w_n$ gets exactly one proposal and never rejects the man that proposes her. So $m_i$ cannot propose to $w_j$ after proposing to $w_n$, since it requires $w_n$ to reject $m_i$ under \DA{} to make that happen. Hence, we reach a contradiction.
\end{proof}
Note that the above lemma claims existence of a woman who is least preferred by every man if the profile satisfies \MP{}. In the proof, we have identified that woman as the woman who receives exactly one proposal in \DA{}.

\begin{lemma}
\label{lemma:struct-maxp-2}
    Suppose, a preference profile $\succ$ satisfies \MP{}. WLOG, $w_n$ be the woman who is every man's last preference in $\succ$, and $m_n$ get matched with $w_n$ in men-proposing \DA{}. Then for each $i \in \{1,\ldots,n-1\}$, $w_i$’s first preference is some $m_j$ ($j \neq n$), and $m_j$’s penultimate preference is $w_i$.
\end{lemma}

\begin{proof}
    From \Cref{lemma:struct-maxp-1}, we know that the 
    woman $w_n$ who is every man's last preference in $\succ$ also receives exactly one proposal in men-proposing \DA{}. By \Cref{lemma:structure-maxp} (as $\succ$ satisfies \MP{}), each woman $w_i \in W\setminus\{w_n\}$ gets proposed by every man in $M$. This implies that she finally gets matched with her most preferred man. Since $m_n$ gets matched with $w_n$, $w_i$'s first preference must be some $m_j$ ($j \neq n$).

    Again using \Cref{lemma:structure-maxp}, $m_j$ proposes to all $(n-1)$ women in $W\setminus\{w_n\}$, and he makes his last proposal to the woman who is finally matched with him, i.e., $w_i$. Since, $m_j$'s least preferred woman is $w_n$, $w_i$ must be $m_j$'s penultimate preference in $\succ$. 
\end{proof}

Using these results, we will now state a set of conditions that are necessary and sufficient for \MP{}. These conditions also identify the additional structure needed for a preference profile in \USM{} to satisfy \MP{}.

\begin{theorem}
    \label{thm:necessary-sufficient-maxp}
    A preference profile $\succ$ satisfies \MP{} (\mMP{}, WLOG) if and only if there exists an ordering $m_1,\ldots,m_n$ of $M$ and an ordering $w_1,\ldots,w_n$ of $W$ satisfying the following three conditions:
    \begin{enumerate}
        \item \label{cond:wn-least} $w_n$ is the least preferred woman for each $m_i \in M, i = 1, \ldots, n$.
        \item \label{cond:wi-penultinate} For each $i \in \{1,\dots,n-1\}$, $w_i$'s first preference is $m_i$, and $m_i$'s penultimate preference is $w_i$.
        \item \label{cond:second-pref} For each $k\in\{1,\dots, n-1\}$, the second preference of $w_k$ is from $\{m_{k+1}, m_{k+2}, \ldots, m_{n}\}$.
    \end{enumerate}
\end{theorem}

Before proving this theorem, we make the following observation on condition~\ref{cond:second-pref}. 
\begin{observation}
\label{obs:digraph}
    Let the second preference of any woman $w_\ell$ be denoted by $s(w_\ell)$. Define $G$ to be the digraph on vertices $\{1,2,\dots,n-1\}$, with an edge from $i$ to $j$ if $s(w_i)=m_j$. Then, there exists an ordering of men and women satisfying condition~\ref{cond:second-pref} of \Cref{thm:necessary-sufficient-maxp} if and only if $G$ is acyclic.
\end{observation}

The above observation is immediate from the insights that (1) an ordering of men and women satisfies condition~\ref{cond:second-pref} of \Cref{thm:necessary-sufficient-maxp} if and only if it gives a topological ordering for $G$, and (2) a directed graph has a topological ordering if and only if it is acyclic. We are now ready to prove \Cref{thm:necessary-sufficient-maxp}.

\begin{proof}[Proof of \Cref{thm:necessary-sufficient-maxp}]    
    $(\Rightarrow)$: Consider a preference profile $\succ$ that satisfies \MP{}. Since $\succ$ satisfies \MP{}, conditions~\ref{cond:wn-least} and \ref{cond:wi-penultinate} of this theorem follow from \Cref{lemma:struct-maxp-1,lemma:struct-maxp-2} respectively. We will prove condition~\ref{cond:second-pref} by showing that the digraph $G$ as defined in \Cref{obs:digraph} is acyclic. Suppose not. Then, $G$ must have at least one directed cycle $C$ involving at least two vertices. Denote the set of vertices in this cycle as $V(C)$. We will show that there exist two different stable matchings, which contradicts that $\succ$ satisfies \MP{} (since \MP{} implies \USM{} by \Cref{thm:maxp-implies-usm}).
    Construct a matching $\mu'$ as follows. For each edge $i,j \in V(C)$ such that a directed edge exists from $i$ to $j$ in $G$, $\mu'(w_{i}) = m_{j}$. For all the remaining women $w_i$, where $i \in N \setminus V(C)$, $\mu'(w_i) = m_i$. Note that $\mu'$ is a stable matching, because of the following reasons.
    \begin{itemize}
        \item None of the women $w_{i}$, where $i \in C$, can form a blocking pair. The only better match the woman $w_{i}$ can get is to be matched with her first preference $m_{i}$ (since she is currently matched to her second preference and condition~\ref{cond:wi-penultinate} says that her top preference is $m_{i}$). But that man $m_{i}$ has $w_{i}$ as the penultimate preference (condition~\ref{cond:wi-penultinate}) and $w_n$ as the last preference (condition~\ref{cond:wn-least}), and is currently matched with neither of them under $\mu'$. So, $m_{i}$ does not find this a profitable deviation.
        \item None of the remaining women $w_i$, where $i \in N \setminus V(C)$, can form a blocking pair either, since $\mu'(w_i) = m_i$, i.e., they have been matched with their most preferred men (condition~\ref{cond:wi-penultinate}), with the exception of $w_n$, who cannot form a blocking pair as she is every man's last preference (condition~\ref{cond:wn-least}).
    \end{itemize}
    However, $\mu(w_i) = m_i$ is also a stable matching, as each $w_i$ gets matched with her most preferred man $m_i$ (except $w_n$ who cannot form a blocking pair due to condition~\ref{cond:wn-least}). Clearly, $\mu \neq \mu'$, since in $\mu'$, at least two women between $1,\ldots,(n-1)$ are matched with their second most preferred men. Thus, we have found two distinct stable matchings $\mu$ and $\mu'$ for $\succ$, which gives us a contradiction to \USM{} (and therefore \MP{}).

    \medskip \noindent
    $(\Leftarrow)$: Consider a preference profile $\succ$ satisfying all three conditions of this theorem. Pick any stable matching $\mu$ on $\succ$. 
    
    First, note that $\mu(w_n)=m_n$, i.e., $w_n$ has to be matched with $m_n$ in every stable matching on $\succ$. This is because if $w_n$ is matched with $m_i \in M \setminus\{m_n\}$ then $(m_i,w_i)$ forms a blocking pair: $m_i$'s least preferred woman is $w_n$ (condition~\ref{cond:wn-least}) and $w_i$'s most preferred man is $m_i$ (condition~\ref{cond:wi-penultinate}).

    We will prove that $\mu(w_i)=m_i$ for all $i$. Suppose not. Let $k$ be largest such that $\mu(w_{k}) \neq m_{k}$. This implies that for all $i \in \{k+1, k+2,\ldots, n\}$, we have $\mu(w_i)=m_i$. Therefore, $w_{k}$ is matched with neither (a)~her first nor (b)~her second preference. This is because, (a)~condition~\ref{cond:wi-penultinate} says that $m_{k}$ is $w_{k}$'s most preferred man, and (b)~ the second preference of $w_k$ i.e. $s(w_k)$ is from $\{m_{k+1}, m_{k+2}, \ldots, m_{n}\}$ (by condition~\ref{cond:second-pref}) but they are matched with $\{w_{k+1}, w_{k+2}, \ldots, w_{n}\}$ respectively (by assumption that $k$ is the largest). But then, $w_{k}$ can form a blocking pair with $m':=s(w_k)$ that is her second preference, as $m'$ has been matched with his least or penultimate preferences, and would prefer $w_{k}$ over $\mu(m')$, and we reach a contradiction.

    Thus $\mu(m_i)=w_i, \forall i \in N,$ is {\em the} unique stable matching for $\succ$, and hence the men-proposed \DA{} algorithm must arrive at this matching. According to this algorithm, each man $m_i$ starts with proposing to his most preferred woman and proposes to the next woman in his preference profile every time he gets rejected, until he reaches his penultimate woman $w_i$ (except for $m_n$, who proposes until he reaches his last preference $w_n$). Each $m_i$ for $i\in\{1,\dots,n-1\}$ proposes $(n-1)$ times, and $m_n$ proposes $n$ times, adding up to a total of  $(n-1)(n-1)+n=n^2-n+1$ proposals. Thus, the preference profile $\succ$ satisfies \MP{}.

    This concludes both directions of the proof.
\end{proof}

\Cref{thm:necessary-sufficient-maxp} gives the necessary and sufficient conditions of \MP{} in the form of three conditions. It is worth asking how critical each of the conditions is. We provide the following three examples to show that each of these conditions is tight.

\begin{example}[Profile $\succ$ violates condition~\ref{cond:wn-least} but satisfies conditions~\ref{cond:wi-penultinate} and \ref{cond:second-pref}]
    Consider the following preference profile $\succ$ for $n=3$.
    \[
    \left(
    \begin{array}{cc}
       m_1: & w_3 \succ w_1 \succ w_2 \\
       m_2: & w_1 \succ w_2 \succ w_3 \\
       m_3: & w_1 \succ w_2 \succ w_3
    \end{array};
    \
    \begin{array}{cc}
       w_1: & m_1 \succ m_2 \succ m_3 \\
       w_2: & m_2 \succ m_3 \succ m_1 \\
       w_3: & m_1 \succ m_2 \succ m_3
    \end{array}
    \right)
    \]
    Observe that $\succ$ satisfies conditions~\ref{cond:wi-penultinate} and~\ref{cond:second-pref} with $\sigma=(2,1)$, but it violates condition~\ref{cond:wn-least}, as $m_1$'s least preferred woman is not $w_3$. Men-proposed \DA{} on $\succ$ yields the matching $\mu=\{(m_1,w_3),(m_2,w_1),(m_3,w_2)\}$, which requires only 4 proposals. If $\succ$ satisfied \mMP{}, it would require $3^2-3+1=7$ proposals. Thus, $\succ$ violates \mMP{}.
\end{example}

\begin{example}[Profile $\succ$ violates condition~\ref{cond:wi-penultinate} but satisfies conditions~\ref{cond:wn-least} and \ref{cond:second-pref}]
    Consider the following preference profile $\succ$ for $n=3$.
    \[
    \left(
    \begin{array}{cc}
       m_1: & w_1 \succ w_2 \succ w_3 \\
       m_2: & w_2 \succ w_1 \succ w_3 \\
       m_3: & w_1 \succ w_2 \succ w_3
    \end{array};
    \
    \begin{array}{cc}
       w_1: & m_1 \succ m_2 \succ m_3 \\
       w_2: & m_2 \succ m_3 \succ m_1 \\
       w_3: & m_1 \succ m_2 \succ m_3
    \end{array}
    \right)
    \]
    Observe that $\succ$ satisfies conditions~\ref{cond:wn-least} and~\ref{cond:second-pref} with $\sigma=(2,1)$, but it violates condition~\ref{cond:wi-penultinate}, as $m_1$ and $m_2$ do not have $w_1$ and $w_2$ respectively as their penultimate preferences. Men-proposed \DA{} on $\succ$ yields the matching $\mu=\{(m_1,w_1),(m_2,w_2),(m_3,w_3)\}$, which requires only 5 proposals. If $\succ$ satisfied \mMP{}, it would require $3^2-3+1=7$ proposals. Thus, $\succ$ violates \mMP{}.
\end{example}

\begin{example}[Profile $\succ$ violates condition~\ref{cond:second-pref} but satisfies conditions~\ref{cond:wn-least} and \ref{cond:wi-penultinate}]
    Consider the following preference profile $\succ$ for $n=3$.
    \[
    \left(
    \begin{array}{cc}
       m_1: & w_2 \succ w_1 \succ w_3 \\
       m_2: & w_1 \succ w_2 \succ w_3 \\
       m_3: & w_1 \succ w_2 \succ w_3
    \end{array};
    \
    \begin{array}{cc}
       w_1: & m_1 \succ m_2 \succ m_3 \\
       w_2: & m_2 \succ m_1 \succ m_3 \\
       w_3: & m_1 \succ m_2 \succ m_3
    \end{array}
    \right)
    \]
    Observe that $\succ$ satisfies conditions~\ref{cond:wn-least} and~\ref{cond:wi-penultinate}, but it violates condition~\ref{cond:second-pref}, as there is no woman $w_{\sigma(1)}$ with $m_3$ as her second most preferred man. Men-proposed \DA{} on $\succ$ yields the matching $\mu=\{(m_1,w_2),(m_2,w_1),(m_3,w_3)\}$, which requires only 5 proposals. If $\succ$ satisfied \mMP{}, it would require $3^2-3+1=7$ proposals. Thus, $\succ$ violates \mMP{}.
\end{example}

 Note that if at any intermediate stage of the men-proposing Gale-Shapley algorithm, $k$ men propose, it can lead to at most $k$ rejections. Only the men who get rejected in a round may propose in the next round. Hence, the following observation is immediate.
\begin{observation}
    \label{fact:at-most-k}
    If there are $k$ men who propose in a particular round, then at most $k$ men (not necessarily the same men) can propose in all subsequent rounds.
\end{observation}
To characterize the distinction between the preference profiles that are \MP{} and \MR{}, we provide the following result that characterizes \MR{} using one additional structural property.

\begin{theorem}
\label{thm:necessary-sufficient-maxr}
    A preference profile $\succ$ satisfies \MR{} if and only if it satisfies the following conditions
    \begin{enumerate}
        \item \label{cond:MP} $\succ$ satisfies \MP{}, and there exists a woman $w_n$ who is the least preferred woman of each man, and
        \item \label{cond:onto} each woman in $W\setminus\{w_n\}$ is a top preference of some man.
    \end{enumerate}
\end{theorem}

\begin{proof}
    $(\Rightarrow):$ Since $\succ$ satisfies \MR{}, it satisfies \MP{} (\Cref{thm:maxr-implies-maxp}) as well, and from \Cref{thm:necessary-sufficient-maxp}, we know that there exists a woman $w_n$ who is the least preferred woman of each man. Hence condition~\ref{cond:MP} is necessary. 

    Also, since $\succ$ satisfies \MR{}, by definition, the number of rounds is $n^2-2n+2$. The first round always has $n$ proposals and since \MR{} $\Rightarrow$ \MP{} (\Cref{thm:maxr-implies-maxp}), the remaining $n^2-n+1-n = n^2-2n+1$ number of proposals have to come in the remaining $n^2-2n+1$ rounds. This implies that each subsequent round must have exactly one proposal. Now, the number of proposals in the second round is equal to the number of men rejected in the first round, which must be $1$. Since all the men propose to some woman amongst the first $(n-1)$ women (\Cref{thm:necessary-sufficient-maxp}), we have that all $(n-1)$ women in $W \setminus \{w_n\}$ must receive at least one proposal in the first round (else, more than one man will be rejected in the first round). This implies that each of the women in $W \setminus \{w_n\}$ must be a top preference of at least one man, which is precisely condition~\ref{cond:onto}.

    $(\Leftarrow):$ Since $\succ$ satisfies \MP{} and every woman in $W\setminus\{w_n\}$ gets a proposal in the first round of the \DA{} algorithm, at most one man can be rejected in that round since exactly one woman gets two proposals. In the second and each subsequent round, we can have at most one proposal (using \Cref{fact:at-most-k}). Since $\succ$ satisfies \MP{}, to get $n^2 - n + 1$ proposals where the first round involves $n$ proposals and every subsequent round involves at most one proposal, we must have $n^2 - 2n + 2$ rounds ($1$ round $\times$ $n$ proposals + remaining $n^2 - 2n + 1$ rounds $\times$ $1$ proposal). Hence $\succ$ satisfies \MR{}.
\end{proof}

Now, a naive way to check if a preference profile $\succ$ satisfies \MP{} (\MR{}) is to run the DA algorithm and check if it achieves the maximum number of proposals (rounds). This would take $\mathcal{O}(n^2)$ time. But, using the characterization of \MP{} (\MR{}), i.e., \Cref{thm:necessary-sufficient-maxp} (\Cref{thm:necessary-sufficient-maxr}), we can do much better. Define the following decision problems \textup{is\MP{}}$(\succ)$ and \textup{is\MR{}}$(\succ)$ as the problems to determine if $\succ$ satisfies \MP{} and \MR{} respectively.

\begin{theorem} 
\label{thm:complexity}
For any preference profile $\succ$ 
    \begin{enumerate}
        \item \textup{is\MP{}}$(\succ)$ can be checked in $\mathcal{O}(n)$.
        \item \textup{is\MR{}}$(\succ)$ can be checked in $\mathcal{O}(n)$.
    \end{enumerate}
\end{theorem}
\begin{proof}
    First, we consider \textup{is\MP{}}$(\succ)$. Clearly, condition~\ref{cond:wn-least} and \ref{cond:wi-penultinate} of Theorem \ref{thm:necessary-sufficient-maxp} can be checked in $\mathcal{O}(n)$ time. Now, we know that whether a directed graph $G(V,E)$ is acyclic can be checked in $\mathcal{O}(|V|+|E|)$ time. Consider that graph $G$,  defined in \Cref{obs:digraph}, has $n-1$ vertices and at most $n-1$ edges. Thus, condition~\ref{cond:second-pref} can also be checked in $\mathcal{O}(n)$ time. Hence, the first part of this theorem is proved. 

    For \textup{is\MR{}}$(\succ)$, note that condition~\ref{cond:onto} of Theorem \ref{thm:necessary-sufficient-maxr} can also be checked in $\mathcal{O}(n)$ time. Hence, combining this with the first part of this theorem, we conclude that \textup{is\MR{}}$(\succ)$ is checkable in $\mathcal{O}(n)$ time.
\end{proof}

\paragraph{Discussions.} 
These results help us understand the \MP{} and \MR{} conditions (and thereby \USM{}) better. 
\begin{enumerate}
    \item The structures look only at partial preferences. The result says we need to know only the top {\em two} preferred alternatives of one side (say women), the bottom {\em two} (top one and bottom {\em two}, for \MR{}) preferred alternatives of the other side (say men), and be agnostic about the preferences at the other positions. Therefore, we can apply this result on domains with partial preferences as long as the preferences at these positions are known. From a practical viewpoint, depending on the applications, such profiles may show up in practice.
    \item From the structure given by \Cref{thm:necessary-sufficient-maxp} (or \Cref{thm:necessary-sufficient-maxr}), it is possible to count what fraction of preference profiles satisfy \MP{} (or \MR{}).
\end{enumerate}

\section{Position of MaxProp in the USM space}

In this section, we analyze the position of \MP{} (and \MR{}) in the class of all preference profiles satisfying USM, relative to known structures contained in this space.

\subsection{\MP{} is disjoint from \SPC{} for $n \geqslant 3$}
\label{sec:disjoint}

Here, we address the relative positions of the \SPC{} and \MP{} classes within the space of \USM{}. We show that these two classes are disjoint when $n\geqslant 3$.

\begin{theorem}
    \label{thm:maxp-spc-disjoint}
    For $n \geqslant 3$, there does not exist any preference profile $\succ \in \mathcal{P}$ that satisfies both \SPC{} and \MP{}.
\end{theorem}

\begin{proof}
    Suppose there exists a preference profile $\succ$ that satisfies both \SPC{} and \MP{}. By definition of \SPC{}, there exists an ordering of men and women such that, for all $i$,
    \begin{enumerate}
        \item man $m_i$ prefers woman $w_i$ over $w_{i+1},w_{i+2},\ldots,w_n$, and
        \item woman $w_i$ prefers man $m_i$ over $m_{i+1},m_{i+2},\ldots,m_n$.
    \end{enumerate}
    Hence, $m_1$ will be proposing to only $w_1$, who will never reject him, as he is her top preference. Thus, $m_1$ makes only one proposal. Since \MP{} holds, we know there are a total of $n^2-n+1$ proposals to be made. Hence, the remaining $n-1$ men make $n^2-n$ proposals, which means each man makes $(n^2-n)/(n-1) = n$ proposals. Since in the men-proposed deferred acceptance algorithm, no man proposes to the same woman twice, each woman has to receive a proposal from all $(n-1)$ men, i.e., each woman receives $\geqslant n-1$ proposals. Thus, there is no woman who receives exactly one proposal, and this contradicts \Cref{fact:single-proposal}. Hence we have the theorem. 
\end{proof}

\paragraph{Discussions.}
This result naturally implies that for $n \geqslant 3$, the classes \SPC{} and \MR{}, \NCC{} and \MP{}, as well as \NCC{} and \MR{} are mutually disjoint (see \Cref{fig:ngreater3} for an illustration). 

\subsection{\mMP{} and \wMP{} are disjoint for $n \geqslant 3$}
\label{sec:men-women-maxp-disjoint}

Here, we show that the \MP{} classes generated by men-proposing and women-proposing \DA{} are disjoint when there are at least {\em three} agents on each side of the market.

\begin{theorem}
    \label{thm:mmp-wmp-disjoint}
    For $n \geqslant 3$, there does not exist any preference profile $\succ \in \mathcal{P}$ that satisfies both \mMP{} and \wMP{}.
\end{theorem}

\begin{proof}
    Suppose there exists a preference profile $\succ \in \mathcal{P}$ satisfying both \mMP{} and \wMP{}. Consider the men-proposing \DA{} algorithm on $\succ$. Since $\succ$ satisfies \mMP{}, by \Cref{cor:maxp-women-get-top}, each $w\in W\setminus\{w_n\}$ is matched with her most preferred man, where $w_n$ is the woman receiving exactly one proposal.
    
    Using \Cref{thm:maxp-implies-usm}, we also know that $\succ$ satisfies USM, i.e., men-proposing \DA{} and women-proposing \DA{} arrive at the same matching. Hence, women-proposing \DA{} on $\succ$ yields a matching in which each $w\in W\setminus\{w_n\}$ is matched with her most preferred man, by making only one proposal. The remaining woman $w_n$ can make at most $n$ proposals. Thus, women-proposing \DA{} on $\succ$ can have at most $1\times (n-1) + n = 2n-1$ proposals.

    Further, $\succ$ satisfies $\wMP{}$, which means women-proposing \DA{} on $\succ$ involves $n^2-n+1$ proposals (\Cref{fact:bound-achievable}). In order for this to happen on $\succ$, it must hold that $n^2-n+1 \leqslant 2n-1$, or $n^2-3n+2\leqslant 0$. However, we know that for $n\geqslant 3, n^2-3n+2>0$. Hence, we have a contradiction.

    Therefore, for $n \geqslant 3$, there is no $\succ\in\mathcal{P}$ satisfying both \mMP{} and \wMP{}.
\end{proof}

\subsection{The curious case of $n=2$}
\label{sec:n=2}

When the number of agents in each side is two, the structure of these spaces looks very different. The classes \MP{} and \MR{} become identical, while \SPC{} and \NCC{} become identical with \USM{}. Quite surprisingly, \MP{} becomes a subset of \SPC{}. Moreover, unlike the $n \geqslant 3$ case, here the \mMP{} and \wMP{} classes overlap partially.

All of the above properties are proved by the theorems that follow, and collecting all these results, the space of these conditions is graphically shown in \Cref{fig:nequal2}. We begin by showing that \MP{} and \MR{} are exactly equivalent for $n=2$.

\begin{theorem}[\MP{} = \MR{}]
    \label{thm:maxp-implies-maxr-n=2}
    For $n=2$, every preference profile $\succ$ satisfying \MP{} also satisfies \MR{}.
\end{theorem}

\begin{proof}
    For $n=2$, the maximum number of rounds is $n^2-2n+2 = 2$ and the maximum number of proposals is $n^2-n+1 = 3$. Now, consider a preference profile $\succ$ satisfying \MP{}. \DA{} on that profile will need to make $3$ proposals. Since round 1 of \DA{} can make at most $2$ proposals (as $n=2$), at least $2$ rounds are required to make $3$ proposals, and thus, $\succ$ satisfies \MR{} as well.
\end{proof}

\citet{clark2006uniqueness} showed that \NCC{} implies \SPC{}. Here, we prove that the converse also holds for $n=2$, making \NCC{} and \SPC{} equivalent.

\begin{theorem}[\SPC{} = \NCC{}]
    \label{thm:spc-implies-ncc-n=2}
    For $n=2$, every preference profile $\succ$ satisfying \SPC{} also satisfies \NCC{}.
\end{theorem}

\begin{proof}
    Consider the preference profile $\succ$ which satisfies \SPC{}. Thus, we have an ordering of men and women (WLOG, assume $(m_1,m_2), (w_1,w_2)$) in which $m_1$ prefers $w_1$ to $w_2$ and $w_1$ prefers $m_1$ to $m_2$. \NCC{} requires that if $w_l \succ_{m_1} w_k$ where $l > k$, then it must imply $w_l \succ_{m_2} w_k$. However, we note that there is no $l > k$ with $w_l \succ_{m_1} w_k$, hence condition~\ref{cond:ncc-1} is vacuously true. It is easy to see that the same is true even for condition~\ref{cond:ncc-2}. Hence, whatever be the preference of $m_2$ and $w_2$, the \NCC{} conditions are always satisfied. This completes the proof.
\end{proof}

It is known that for $n=2$, \SPC{} also becomes necessary for \USM{}~\citep{eeckhout2000uniqueness}. Here we provide a direct proof of this result.

\begin{theorem}[\SPC{} = \USM{}]
    \label{thm:usm-implies-spc-n=2}
    For $n=2$, preference profile $\succ$ satisfies \SPC{} if and only if it is in \USM{}.
\end{theorem}

\begin{proof}
    Note that the `only if' direction comes directly from \citet{eeckhout2000uniqueness}, since the proof holds even for $n=2$. Hence, we only show the `if' direction of this result.

    We will show that if a profile $\succ$ does not satisfy \SPC{} then it cannot belong to \USM{}. Note that, for \SPC{} to be violated, it is necessary that there does not exist a pair of man and woman who rank each other as their first preference. To make this happen, for $n=2$, both men cannot have the same woman as their first preference, and both women should also have the man who {\em does not} rank her at the top as her first preference. Hence, the only two possible preference profiles are
    \[ 
    \left(
    \begin{array}{cc}
       m_1: & w_1 \succ w_2 \\
       m_2: & w_2 \succ w_1 
    \end{array};
    \
    \begin{array}{cc}
       w_1: & m_2 \succ m_1 \\
       w_2: & m_1 \succ m_2
    \end{array}
    \right)
    \text{ or }
    \left(
    \begin{array}{cc}
       m_1: & w_2 \succ w_1 \\
       m_2: & w_1 \succ w_2
    \end{array};
    \begin{array}{cc}
       w_1: & m_1 \succ m_2 \\
       w_2: & m_2 \succ m_1 
    \end{array}
    \right).
    \]
    In both the profiles, the men-optimal \DA{} yields a different matching that the women-optimal \DA{}. Hence, this profile does not belong to \USM{}. This concludes the proof.
\end{proof}

Note that \citet{eeckhout2000uniqueness} claims \SPC{} to be necessary for \USM{} even for $n=3$, which is not true. As we show in \Cref{ex:usm-not-maxp-not-spc}, there are profiles for $n=3$ that are not \SPC{} but admit a unique stable matching. Next, we prove that \MP{} is a proper subset of \SPC{} even for the case of $n=2$.

\begin{theorem}[\MP{} $\subset$ \SPC{}]
    \label{thm:maxp-implies-spc-n=2}
    For $n=2$, every preference profile $\succ$ satisfying \MP{} also satisfies \SPC{}.
\end{theorem}

\begin{proof}
    Observe that for $n=2$, if both men have the same top women in their preference list, then it is sufficient to claim that the profile is \SPC{}. This is because, the woman (say $w_1$) who is this top choice of both the men has exactly one man as her top choice (say $m_1$). Then it is easy to see that the order $(m_1,m_2), (w_1,w_2)$ is the \SPC{} satisfying order.

    Now, let a profile $\succ$ satisfy \MP{}. For $n=2$, it implies that the men should make $2^2-2+1 = 3$ proposals. If their top preferences were different women, then \DA{} would complete in round 1 with $2$ proposals. Hence, it is necessary to have the same woman as the top preference of both men for $\succ$ to be in \MP{}. With our previous observation, we conclude that $\succ$ also satisfies \SPC{}.
\end{proof}

The converse of the above result is not true. Indeed, \MP{} is a strict subset of \SPC{} as the following example shows.

\begin{example}[\SPC{} but not \MP{} for $n=2$]
    \label{ex:spc-not-mp-n=2}
    Consider the following preference profile.
    \[
    \left(
    \begin{array}{cc}
       m_1: & w_1 \succ w_2 \\
       m_2: & w_2 \succ w_1 
    \end{array};
    \
    \begin{array}{cc}
       w_1: & m_1 \succ m_2 \\
       w_2: & m_2 \succ m_1
    \end{array}
    \right)
    \]
    It is easy to see that \SPC{} is satisfied on this profile with the order being $(m_1,m_2),(w_1,w_2)$. However, the number of proposals in men-proposing \DA{} is $2$ while \MP{} requires this to be $2^2-2+1 = 3$. Hence, this profile does not satisfy \MP{}.
\end{example}

\paragraph{Relative structures of \mMP{} and \wMP{}.}
Unlike the $n \geqslant 3$ case, here these two classes overlap partially. 

\begin{theorem}
    \label{thm:partial-overlap-n=2}
    For $n=2$, a preference profile $\succ \in \mathcal{P}$ 
    \begin{enumerate}
        \item \label{overlap-n=2-part1} satisfies \mMP{} iff both men have the same woman as their top preference, and
        \item \label{overlap-n=2-part2} satisfies both \mMP{} and \wMP{} iff in addition to the above condition both women also have the same man as their top preference.
    \end{enumerate}
\end{theorem}

\begin{proof}
    Part~\ref{overlap-n=2-part1}: Consider the `if' direction. If both men have the same woman as the top preference in $\succ$, then in first round of men-proposing \DA{}, two proposals will be made and one of them will be rejected who will propose in the next round. Since the maximum number of proposals for $n=2$ is $2^2 -2 + 1 = 3$, this will lead to \mMP{}. For the `only if' direction, suppose the two men do not have the same woman as their top preference. Then the men-proposing \DA{} will get over in one round with two proposals, and hence will not belong to \mMP{}.

    \medskip \noindent
    Part~\ref{overlap-n=2-part2}: Now we know that the \mMP{} class contains only those profiles where the men have the same woman as their top preference. In addition, if we also need the profile $\succ$ to be \wMP{}, then using the women-equivalent condition of Part~\ref{overlap-n=2-part1}, we get that it is equivalent to both women also having the same man as their top preference. Therefore, the necessary and sufficient condition for a preference profile to be both \mMP{} and \wMP{} is that both men have the same woman as their top preference and both women also have the same man as their top preference.
\end{proof}


\section{Conclusions and Future Work}
\label{sec:concl}

We considered the \USM{} problem from a \citet{gale1962college} {\em deferred acceptance algorithmic} perspective. The properties like \MP{} and \MR{} that count the number of proposals and rounds respectively in this algorithm yield novel insights into the structure of \USM{}. Both the \MP{} and \MR{} properties are computationally easy to verify (\Cref{thm:necessary-sufficient-maxp}) without invoking the \DA{} algorithm.
In addition, these conditions carve out a different and unexplored sub-space of \USM{} (see \Cref{fig:contrib_illus}). The variation of these spaces for $n=2$ and $n \geqslant 3$ is interesting.

As a future plan, we would like to see if any algorithmic property (of not necessarily \DA{}) can explain the whole of the \USM{} class and if there exists an efficient (better than \DA{}) algorithm that can identify \USM{}. We would also like to explore the generalization of our results to the setting of many-to-one matchings.

\section*{Acknowledgments}

SN acknowledges the support of a MATRICS grant (MTR/2021/000367) and two Core Research Grants (CRG/2022/009169 and CRG/2023/001442) from SERB, Govt. of India. He also acknowledges the support of a SBI Foundation research grant (DO/2023-SBIF002-005), a TCAAI grant (DO/2021-TCAI002-009), a TCS research grant (MOU/CS/10001981-1/22-23), and an Amazon research grant (PO: 2D-09926960).

\bibliographystyle{named}
\bibliography{References}


\appendix

\section{\MP{} = \MR{} for $n = 3$}
\label{app:mp=mr-n=3}

As remarked in \Cref{sec:maxR-maxP}, \MP{} and \MR{} are equivalent in the case of $n=3$. We prove this by the following lemma.

\begin{lemma}
 \label{lemma:mp-mr-n=3}
    For $n=3$, if preference profile $\succ$ satisfies \MP{}, then it also satisfies \MR{}.
\end{lemma}
\begin{proof}
    WLOG, suppose $\succ$ satisfies \mMP{}. Consider the ordering over men and women implied by \Cref{thm:necessary-sufficient-maxp}.
    
    By condition~\ref{cond:wn-least} of \Cref{thm:necessary-sufficient-maxp}, the third preference of every man is $w_3$. Moreover, by condition~\ref{cond:wi-penultinate} of \Cref{thm:necessary-sufficient-maxp}, the second preferences of $m_1$ and $m_2$ are $w_1$ and $w_2$ respectively. This leads us to conclude that the top preferences of $m_1$ and $m_2$ are $w_2$ and $w_1$ respectively.

    Thus, each woman in $W\setminus\{w_n\}=\{w_1,w_2\}$ is the top preference of some man, where $w_n=w_3$ is each man's last preference. This is exactly condition~\ref{cond:onto} of \Cref{thm:necessary-sufficient-maxr}, and we already have condition~\ref{cond:MP} since $\succ$ satisfies \MP{}. Hence, by \Cref{thm:necessary-sufficient-maxr}, $\succ$ satisfies \MR{}.
\end{proof}

Together with the fact that \MR{} $\implies$ \MP{} (\Cref{thm:maxr-implies-maxp}), we conclude that the conditions \MP{} and \MR{} are equivalent for $n=3$.

\end{document}